\documentclass[copyright,creativecommons]{eptcs}
\usepackage{prelude}
\providecommand{\event}{TTC 2013}
\providecommand{\authorrunning}{W. Smid \& A. Rensink}
\providecommand{\titlerunning}{Class Diagram Restucturing with \GROOVE}
\begin{document}
\title{Class Diagram Restucturing with \GROOVE}
\author{Wietse Smid 
\institute{University of Twente \\ The Netherlands \\ \email{w.b.t.smid@student.utwente.nl}}
\and Arend Rensink
\institute{University of Twente \\ The Netherlands \\ \email{arend.rensink@utwente.nl}}
}
\maketitle
\begin{abstract}
  This paper describes the \GROOVE solution to the ``Class Diagram
  Restructuring'' case study of the Tool Transformation Contest 2013. We show
  that the visual rule formalism enables the required restructuring to be
  formulated in a very concise manner. Moreover, the \GROOVE functionality for
  state space exploration allows checking confluence. Performance-wise,
  however, the solution does not scale well.
\end{abstract}

%%% Local Variables: 
%%% mode: latex
%%% TeX-master: "paper"
%%% End: 

\section{Introduction}
 
This paper describes the \GROOVE solution to the ``Class Diagram
Restructuring'' case study of the Tool Transformation Contest 2013 \cite{CDR}.

\GROOVE \cite{Rensink2003,STTT} is designed to support state space exploration
and model checking of dynamically evolving systems; however, as we will show
in this contribution, it can also be used to demonstrate graph transformations
as a basis for model transformation.  \GROOVE's main usage in this solution
(and indeed the core feature of \GROOVE) is to formulate the required changes
as transformation rules and explore all possible ways to apply those rules to
a given graph, representing the initial class diagram.

\medskip\noindent\textbf{Graphs and rules.}
In \GROOVE, graphs are defined as nodes and directed edges with labels. Nodes
have a special type of labels: the type label defines the type for a node,
this is used in conjunction with a type graph system that preserves type
safety. Furthermore, attributes are supported in the form of directed edges to
value nodes.\cite{Kastenberg2005}

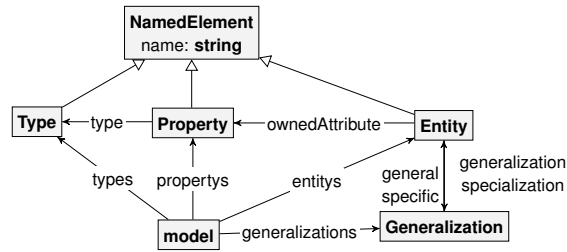
\begin{figure}
  \begin{center}
  \scalebox{.8}{% To use this figure in your LaTeX document
% import the package groove/resources/groove2tikz.sty
%
\begin{tikzpicture}[scale=\tikzscale]
\node[type_node] (n2) at (4.265, -1.075) {\ml{\textbf{Entity}}};
\node[type_node] (n3) at (2.170, -0.305) {\ml{\textbf{NamedElement}\\name: \textbf{string}}};
\node[type_node] (n5) at (0.885, -1.045) {\ml{\textbf{Type}}};
\node[type_node] (n8) at (4.270, -1.915) {\ml{\textbf{Generalization}}};
\node[type_node] (n7) at (2.145, -1.985) {\ml{\textbf{model}}};
\node[type_node] (n6) at (2.180, -1.055) {\ml{\textbf{Property}}};

\path[basic_edge](n2.south -| 4.270, -1.915) --  (n8) 
node[lab] at (4.840, -1.472) {\ml{generalization\\specialization}};
\path[basic_edge](n2.west |- 2.180, -1.055) -- node[lab] {\ml{ownedAttribute}} (n6) ;
\path[basic_edge](n8.north -| 4.265, -1.075) --  (n2) 
node[lab] at (3.988, -1.562) {\ml{general\\specific}};
\path[subtype_edge] (n2)  --  (n3) ;
\path[basic_edge] (n7)  -- node[lab] {\ml{types}} (n5) ;
\path[basic_edge](n7.north -| 2.180, -1.055) -- node[lab] {\ml{propertys}} (n6) ;
\path[basic_edge] (n7)  -- node[lab] {\ml{entitys}} (n2) ;
\path[subtype_edge] (n5)  --  (n3) ;
\path[basic_edge](n6.west |- 0.885, -1.045) -- node[lab] {\ml{type}} (n5) ;
\path[subtype_edge](n6.north -| 2.170, -0.305) --  (n3) ;
\path[basic_edge] (n7)  -- node[lab] {\ml{generalizations}} (n8) ;
\end{tikzpicture}}
  \caption{\GROOVE type graph created from the \ECORE metamodel}
  \label{fig:ecore}
  \end{center}
  \end{figure}

Rules enable the manipulation of graphs. Main rule features are:

\begin{description}[nosep]
\item[Readers:] Edges and nodes that must be present for a rule to
  be applicable, and are unaffected by the transformation. Readers are
  visually distinguished by black, continuous lines.

\item[Erasers:] Edges and nodes that (like readers) must be present
  for a rule to be applicable, but are deleted when the rule
  is applied. Erasers are visually distinguihsed by dashed blue lines.

\item[Creators:] Edges and nodes that are created when the rule is
  applied. Creators are visually distinguished by fat green lines.

\item[Embargoes:] Edges and nodes that are forbidden; i.e., when
  present they prevent a rule from being applicable. Embargoes are visually
  distinguished by fat, dotted red lines.

\item[Quantification:] Universal and Existential quantification are possible
  via ``$\forall$'' and '$\exists$" nodes. Other nodes can be bound to these
  quantifiers using edges labeled ``@''. Quantifiers can also be nested using
  edges labeled ``in''. For more details see \cite{RensinkKup2009}.
\end{description}
In the Class Diagram Restructing case study, the start graph was provided by
an \ECORE model. Since recently (see \cite{Bruintjes2012}), \GROOVE can import
and export \ECORE models. When importing, the user can make choices regarding
the representation of certain \ECORE features that have no direct \GROOVE
counterpart. For instance, the models in this case study have opposing edges
between \type{Generalization} and \type{Entity} (e.g., \lab{generalization}
vs.\lab{specific}); in the chosen \GROOVE representation, these are modelled
by separate unidirectional edges. The actual type graph is shown in Fig.~\ref{fig:ecore}. Another
possible choice would have been to introduce an intermediate node for each
pair of concrete opposing edges; that blows up the model but retains more
information, viz.\ the fact that the edges actually belong together.

To use \GROOVE to actually carry out the transformations of this case study,
the following steps are required:
\begin{enumerate}[noitemsep]
\item Load the rule system into the Simulator (the GUI component of \GROOVE);
\item Import the \texttt{.xmi} file of the input model using the \ECORE importer;
\item Explore the state space, either manually (step by step) or automatically
  (completely);
\item Export the final state of the exploration using the \ECORE exporter.
\end{enumerate}
Alternatively, the exploration can be run from the command line using the
Generator (the headless component of \GROOVE).

%%% Local Variables: 
%%% mode: latex
%%% TeX-master: "paper"
%%% End: 

\section{Transformation Steps and Rule Applications}
\label{sec:trans}

The case study calls for three distinct types of transformation steps; these
should be executed to transform the given graph to its final state, viz.\ the
restructured class diagram.

\paragraph{Step 1: Pull Up common attributes of all direct subclasses}

The first transformation step is by far the easiest and therefore has the
simplest translation: see Fig.~\ref{fig:PullUpDirect}. Called \PullUpDirect,
it searches for all \type{Property}s of which the containing \type{Entity} has
a given \type{Entity} as its superclass. It then checks if all the
\type{Entity}s with that superclass have the same \type{Property} in terms of
\lab{type} and \lab{name}. The number of subclasses that have the
\type{Property} must be larger than 1.  The \type{Property}s are removed from
the independent classes and a \type{Property} with the same \lab{type} and
\lab{name} is created for the superclass. Note that the \lab{in}-labelled
(dotted) edge from the $\exists$-node to the $\forall$-node signifies nesting
of the quantifiers: \emph{for all} \type{Generalization}+\type{Entity}-pairs
\emph{there exists} a \type{Property}.

  \begin{figure}
  \begin{center}
  \scalebox{.8}{% To use this figure in your LaTeX document
% import the package groove/resources/groove2tikz.sty
%
\begin{tikzpicture}[scale=\tikzscale]
\node[basic_node] (n0) at (2.035, -1.380) {\ml{\textbf{Generalization}}};
\node[basic_node] (n1) at (1.960, -2.130) {\ml{\textbf{Entity}}};
\node[basic_node] (n2) at (2.770, -0.580) {\ml{\textbf{Entity}}};
\node[nesting_node] (n3) at (0.750, -1.810) {\ml{$\forall$\\count $>$ 1}};
\node[eraser_node] (n4) at (1.990, -3.070) {\ml{\textit{b} : \textbf{Property}\\a.name == name}};
\node[basic_node] (n5) at (3.580, -3.750) {\ml{\textbf{Type}}};
\node[creator_node] (n6) at (4.890, -1.450) {\ml{\textbf{Property}\\{\color{creator_c}name := a.name}}};
\node[basic_node] (n8) at (3.575, -3.030) {\ml{\textit{a} : \textbf{Property}}};
\node[basic_node] (n9) at (3.580, -2.150) {\ml{\textbf{Entity}}};
\node[nesting_node] (n10) at (0.720, -3.040) {\ml{$\exists$}};
\node[basic_node] (n7) at (3.625, -1.380) {\ml{\textbf{Generalization}}};
\node[basic_node] (n11) at (5.095, -0.580) {\ml{\textbf{model}}};

\path[eraser_edge](n4.south -| 2.010, -3.770) -- (2.010, -3.770) -- (n5.west |- 2.010, -3.770)
node[lab] at (2.438, -3.770) {\ml{type}};
\path[eraser_edge](n1.south -| 1.990, -3.070) --  (n4) 
node[lab] at (1.975, -2.588) {\ml{ownedAttribute}};
\path[nesting_edge] (n0)  -- node[lab] {\ml{@}} (n3) ;
\path[basic_edge] (n7)  -- node[lab] {\ml{general}} (n2) ;
\path[basic_edge](n7.south -| 3.580, -2.150) -- node[lab] {\ml{specific}} (n9) ;
\path[creator_edge](n6.south -| 4.870, -3.770) -- (4.870, -3.770) -- (n5.east |- 4.870, -3.770)
node[lab] at (4.870, -3.258) {\ml{type}};
\path[nesting_edge] (n1)  -- node[lab] {\ml{@}} (n3) ;
\path[basic_edge](n9.south -| 3.575, -3.030) -- node[lab] {\ml{ownedAttribute}} (n8) ;
\path[basic_edge](n8.south -| 3.580, -3.750) -- node[lab] {\ml{type}} (n5) ;
\path[basic_edge](n0.south -| 1.960, -2.130) -- node[lab] {\ml{specific}} (n1) ;
\path[basic_edge] (n0)  -- node[lab] {\ml{general}} (n2) ;
\path[nesting_edge](n10.north -| 0.750, -1.810) -- node[lab] {\ml{in}} (n3) ;
\path[nesting_edge](n4.west |- 0.720, -3.040) -- node[lab] {\ml{@}} (n10) ;
\path[creator_edge] (n11)  -- node[lab] {\ml{propertys}}(n6.north -| 5.095, -0.580);
\path[creator_edge](n2.east |- 4.580, -0.600) -- (4.580, -0.600) -- (n6.north -| 4.580, -0.600)
node[lab] at (3.870, -0.590) {\ml{ownedAttribute}};
\end{tikzpicture}}
  \caption{\PullUpDirect rule}
  \label{fig:PullUpDirect}
  \end{center}
  \end{figure}

\paragraph{Step 2: Create subclasses for duplicated attributes}

To execute the creation of a subclass for duplicated attributes we divide the
work into three consecutive rules. We guarantee the consecutive application of
these sub steps by using embargoes in the rules on the temporary nodes that are
created during previous steps.

First we apply the rule \SelectSubClass (Fig.~\ref{fig:SubClass}). Similar to
\PullUpDirect, it matches \type{Entity}s that have a superclass and a
\type{Property}. It then matches all the \type{Entity}s that fall under the
superclass and have a \type{Property} with the same \lab{name} and
\lab{type}. This explicitly does not mean that all the subclasses need this
property. For this match it creates a \type{SelectSub} node for the selected
superclass entity, and it links to all the names of the variables that are a
candidate for upwards movement in the hierarchy.

\begin{figure}
\centering
\scalebox{.8}{% To use this figure in your LaTeX document
% import the package groove/resources/groove2tikz.sty
%
\begin{tikzpicture}[scale=\tikzscale]
\node[basic_node] (n13) at (3.270, -1.345) {\ml{\textbf{Generalization}}};
\node[basic_node] (n12) at (4.965, -1.335) {\ml{\textbf{Entity}}};
\node[embargo_node] (n10) at (3.195, -0.495) {\ml{\textbf{SelectSub}}};
\node[data_node] (n9) at (4.970, -3.375) {\ml{\textbf{string}}};
\node[creator_node] (n3) at (4.975, -2.165) {\ml{\textbf{SelectSub}}};
\node[nesting_node] (n6) at (4.320, -2.710) {\ml{$\forall^{>0}$}};
\node[basic_node] (n2) at (3.290, -3.815) {\ml{\textit{a} : \textbf{Property}}};
\node[basic_node] (n0) at (3.275, -2.695) {\ml{\textbf{Entity}}};
\node[basic_node] (n7) at (5.430, -3.835) {\ml{\textit{type} : \textbf{Type}}};
\node[nesting_node] (n8) at (6.070, -2.700) {\ml{$\forall$\\count $>$ 1}};
\node[basic_node] (n5) at (7.235, -3.820) {\ml{\textit{b} : \textbf{Property}\\a.name == b.name}};
\node[basic_node] (n4) at (7.205, -2.675) {\ml{\textbf{Entity}}};
\node[basic_node] (n1) at (7.230, -1.355) {\ml{\textbf{Generalization}}};
\node[embargo_node] (n11) at (3.530, -0.925) {\ml{\textbf{SelectedSubProperty}}};

\path[basic_edge](n1.south -| 7.205, -2.675) -- node[lab] {\ml{specific}} (n4) ;
\path[creator_edge](n3.south -| 4.970, -3.375) --  (n9) 
node[lab] at (4.960, -2.429) {\ml{name}};
\path[basic_edge](n2.east |- 5.430, -3.835) -- node[lab] {\ml{type}} (n7) ;
\path[basic_edge](n13.east |- 4.965, -1.335) -- node[lab] {\ml{general}} (n12) ;
\path[nesting_edge] (n13)  -- node[lab] {\ml{@}} (n6) ;
\path[nesting_edge] (n7)  -- node[lab] {\ml{@}} (n8) ;
\path[nesting_edge](n4.west |- 6.070, -2.700) -- node[lab] {\ml{@}} (n8) ;
\path[basic_edge](n5.west |- 5.430, -3.835) -- node[lab] {\ml{type}} (n7) ;
\path[creator_edge](n3.north -| 4.965, -1.335) -- node[lab] {\ml{super}} (n12) ;
\path[basic_edge] (n2)  -- node[lab] {\ml{name}} (n9) ;
\path[nesting_edge] (n2)  -- node[lab] {\ml{@}} (n6) ;
\path[nesting_edge](n0.east |- 4.320, -2.710) -- node[lab] {\ml{@}} (n6) ;
\path[basic_edge](n13.south -| 3.275, -2.695) -- node[lab] {\ml{specific}} (n0) ;
\path[basic_edge](n4.south -| 7.235, -3.820) -- node[lab] {\ml{ownedAttribute}} (n5) ;
\path[basic_edge](n0.south -| 3.290, -3.815) -- node[lab] {\ml{ownedAttribute}} (n2) ;
\path[nesting_edge] (n9)  -- node[lab] {\ml{@}} (n8) ;
\path[nesting_edge](n8.west |- 4.320, -2.710) -- node[lab] {\ml{in}} (n6) ;
\path[nesting_edge] (n1)  -- node[lab] {\ml{@}} (n8) ;
\path[basic_edge](n1.west |- 4.965, -1.335) -- node[lab] {\ml{general}} (n12) ;
\path[nesting_edge] (n5)  -- node[lab] {\ml{@}} (n8) ;
\end{tikzpicture}} %\qquad
\scalebox{.8}{% To use this figure in your LaTeX document
% import the package groove/resources/groove2tikz.sty
%
\begin{tikzpicture}[scale=\tikzscale]
\node[basic_node] (n0) at (2.490, -2.465) {\ml{\textbf{Generalization}}};
\node[eraser_node] (n1) at (4.180, -1.595) {\ml{\textit{set} : \textbf{SelectSub}}};
\node[basic_node] (n2) at (2.445, -3.365) {\ml{\textbf{Entity}}};
\node[basic_node] (n3) at (2.450, -4.325) {\ml{\textbf{Property}}};
\node[data_node] (n4) at (4.190, -4.315) {\ml{\textbf{string}}};
\node[creator_node] (n5) at (4.200, -5.220) {\ml{\textbf{SelectedSubProperty}\\{\color{creator_c}count := collect.count}}};
\node[nesting_node] (n6) at (1.140, -5.955) {\ml{$\forall$}};
\node[nesting_node] (n7) at (1.085, -3.800) {\ml{\textit{collect} : $\forall$\\count $>$ 1}};
\node[basic_node] (n8) at (2.470, -5.325) {\ml{\textit{type} : \textbf{Type}}};
\node[basic_node] (n9) at (2.445, -1.575) {\ml{\textbf{Entity}}};

\path[nesting_edge] (n0)  -- node[lab] {\ml{@}} (n7) ;
\path[nesting_edge](n5.south -| 4.200, -5.900) -- (4.200, -5.900) -- (n6.east |- 4.200, -5.900)
node[lab] at (4.190, -5.750) {\ml{@}};
\path[nesting_edge](n7.south -| 1.130, -4.910) -- (1.130, -4.910) -- (n6.north -| 1.130, -4.910)
node[lab] at (1.129, -4.884) {\ml{in}};
\path[nesting_edge] (n3)  -- node[lab] {\ml{@}} (n7) ;
\path[creator_edge] (n3)  -- node[lab] {\ml{in}} (n5) ;
\path[basic_edge](n3.south -| 2.470, -5.325) -- node[lab] {\ml{type}} (n8) ;
\path[basic_edge](n0.south -| 2.445, -3.365) -- node[lab] {\ml{specific}} (n2) ;
\path[eraser_edge](n1.south -| 4.190, -4.315) -- node[lab] {\ml{name}} (n4) ;
\path[basic_edge](n2.south -| 2.450, -4.325) -- node[lab] {\ml{ownedAttribute}} (n3) ;
\path[basic_edge](n3.east |- 4.190, -4.315) --  (n4) 
node[lab] at (3.019, -4.317) {\ml{name}};
\path[creator_edge](n5.west |- 2.470, -5.325) -- node[lab] {\ml{type}} (n8) ;
\path[nesting_edge] (n8)  -- node[lab] {\ml{@}} (n7) ;
\path[creator_edge](n5.north -| 4.190, -4.315) -- node[lab] {\ml{name}} (n4) ;
\path[eraser_edge](n1.west |- 2.445, -1.575) -- node[lab] {\ml{super}} (n9) ;
\path[basic_edge](n0.north -| 2.445, -1.575) -- node[lab] {\ml{general}} (n9) ;
\path[nesting_edge] (n2)  -- node[lab] {\ml{@}} (n7) ;
\path[nesting_edge](n4.east |- 5.110, -4.330) -- (5.110, -4.330) -- (5.100, -6.000) -- (n6.east |- 5.100, -6.000)
node[lab] at (5.096, -5.047) {\ml{@}};
\end{tikzpicture}}
\caption{Rules for \SelectSubClass and \CountSubClass}
\label{fig:SubClass}
\end{figure}

Secondly we count for each of these candidates the number of occurrences, as
we want to execute only the largest case. The rule \CountSubClass does exactly
this. For each of the variables linked in the \type{SelectSub} node it takes
the linked superclass \type{Entity} node and selects all the subclasses that
have a variable of the name. For each of the linked names a
\type{SelectedSubProperty} node is created that is linked to the properties,
the type, and the name of the property. Furthermore the universal quantifier
that is used to select the entities and properties for each of these linked
names has a variable count, which records the number of matched items.

Finally we can then create the new class using a rule \CreateSubClass (omitted
here for lack of space).
%(Fig.~\ref{fig:CreateSubClass}). 
It matches the \type{SelectedSubProperty}
with the greatest number of elements. The generalizations that link the
\type{Entity}s to the superclass are removed, as well as the properties
themselves. A new \type{Entity}, subclass to the original superclass and
superclass to the original subclasses is inserted, having a property of the
same name and type as the matched property in \type{SelectedSubProperty}.
%
%\rulefig{CreateSubClass}{\CreateSubClass rule}

\paragraph{Step 3: Create root class for duplicated attributes}

Step 3 is actually a special case of step 2; only here we do not match on a
common superclass, but on a common shared property, namely the absence of a
superclass. The rules are therefore changed a bit, but the effective procedure
matches that of step 2: first a root class is selected, then a count is
performed for all suitable \type{Properties}, and finally the one with the
maximum count is selected and a new root class is created for it.

\paragraph{Extension to multiple inheritance}

Because \GROOVE is a general-purpose transformation tool, there is no specific
reason in terms of graph transformations why multiple inheritance is
impossible. There are of course a few things to consider.

Representing multiple inheritance is easy enough; the ecore type graph could
allow for more than one general or specific to a generalization node. Trying
to find if there is a shared superclass is then a case of finding a match
where these edges exist.

Furthermore, a situation can be thought of where all subclasses of a class
either have an identical \type{property}, or have a superclass with that property,
with the added constraint that that superclass cannot have subclasses outside
of the original superclass. In this way, rules for multiple inheritance can
get quite complicated; however, to create them is still eminently possible.

The visual interface that \GROOVE provides is a useful tool for this job. All
in all, constructing the rules for multiple inheritance is inherently
difficult, but certainly possible with \GROOVE.

%%% Local Variables: 
%%% mode: latex
%%% TeX-master: "paper"
%%% End: 

\section{Discussion and conclusion}

\GROOVE has its pros and cons as a tool for this case study. On the upside,
\GROOVE provides a few interesting features that where not required in the
scope of the case description. The most important feature is the total state
space exploration, this allows us to visualise things like confluence of a
rule system as well as how a refactoring was reached.

Another positive point is the ease of use of the \GROOVE tool. Because of its
visual interface, designing rules and understanding them is relatively
easy. Furthermore the rule application highlights the matched nodes and edges,
allowing for direct feedback on created rules.

\medskip\noindent\textbf{Performance.}
These advantages come at a price. Because \GROOVE was created for a wider
application area and for a different core functionality, even medium sized
models pose a performance problem for \GROOVE: the amount of resources
needed to complete these examples outstrips what the Java VM can handle
on a regular sized computer. Provided with 10G of memory on a 64-bit machine
(Intel i7-2600 CPU @ 3.40GHz), the performance on the example cases is
summarised in the following table:
\begin{center}
\begin{tabular}{lcccc}
\bf Name & \bf Size & \bf \#States & \bf Time (s) \\
testcase1 & 14   &  3 & $<1$ \\
testcase2 & 18   &  9 & $<1$ \\
testcase3 & 5500 & 13 & 212 \\
testcase2\_1000 & 10000 & Out of memory
\end{tabular}
\end{center}
The size is the combined number of classes and attributes; the number of
states is the set of consecutive graphs that were computed in the course of
the exploration. Testcase 3 involves a very large combined refactoring: 10
attributes are pulled up from 500 classes. Testcase 2\_1000 is a 1000-fold
copy of testcase 2. Clearly, the performence of these larger cases is not
practical.

\medskip\noindent\textbf{Rule scheduling.}
The three step process for transformations 2 and 3 (see Sect.~\ref{sec:trans})
has an inbuilt sequence that is enforced by the embargoes specifying the
absence of the \type{Selected*} nodes. This is a rather crude way of
scheduling rule applications. \GROOVE does provide another way to handle this,
through control programs that explicitly specify that \SelectSubClass,
\CountSubClass and \CreateSubClass must always be performed in that order and
never be interrupted by other rules. This feature has not been used here.

\medskip\noindent\textbf{Case ambiguities.}
The Case Description leaves a few questions open. Firstly in the
application of Rule~2 and Rule~3 it is posed that the first application of
these rules should be on the property with the highest number of
occurrences. There are however situations where there are two options of equal
size which are not isomorphic, thus creating two final transformed graphs
that are not confluent.

The other is the ambiguity in the naming of newly created
\type{Entity}s. Although \GROOVE has no real need for the names of
\type{Entity}s --- it can simply work with nameless entities because it only
works on nodes and their edges --- there is still the need for giving
\type{Entity}s distinct and useful names. This is not a problem when exporting
the graph from \GROOVE to an \texttt{xmi} file, because \GROOVE uses the
internal IDs of the nodes for naming.

\medskip\noindent\textbf{Conclusion.}
The provided solution can transform Class Diagrams according to the given
transformation rules. Even though the solution lacks the performance to solve
the largest test cases, there are other benefits not required from the case
description. Confluence can be shown to hold in cases where the entire state
space can be calculated. Furthermore the \GROOVE tool has a high
ease-of-use factor, and can be used for many more similar problems.

%%% Local Variables: 
%%% mode: latex
%%% TeX-master: "paper"
%%% End: 

\bibliographystyle{eptcs}
\bibliography{paper}

%%% Local Variables: 
%%% mode: latex
%%% TeX-master: "paper"
%%% End: 

\end{document}